\documentstyle[prd,aps,epsfig]{revtex} 

\begin{document} 

\title{%
\vskip-6pt \hfill {\rm\normalsize UCLA/99/TEP/19} \\ 
\vskip-6pt \hfill {\rm\normalsize MPI-PhT/99-22} \\ 
\vskip-6pt \hfill {\rm\normalsize May 1999} \\
\vskip-9pt~\\
Prompt atmospheric neutrinos and muons:\\ 
dependence on the gluon distribution function
}
 
\author{Graciela Gelmini\rlap{,}{$^{1}$} Paolo Gondolo\rlap{,}{$^{2}$} 
Gabriele Varieschi$^{1}$} 
 
\address{
~\\
${}^{1}$ Dept. of Physics and Astronomy, UCLA (University of 
California, Los Angeles)\\
405 Hilgard Ave., Los Angeles CA 90095, USA\\
{\rm gelmini,variesch@physics.ucla.edu}
\\~\\
${}^{2}$ Max-Planck-Institut f\"ur Physik (Werner-Heisenberg-Institut) 
\\
F\"ohringer Ring 6, 80805 M\"unchen, Germany\\
{\rm gondolo@mppmu.mpg.de}
} 
 
\maketitle

\begin{abstract} 
  We compute the next-to-leading order QCD predictions for the vertical flux of
  atmospheric muons and neutrinos from decays of charmed particles, for
  different PDF's (MRS-R1, MRS-R2, CTEQ-4M and MRST) and different
  extrapolations of these at small partonic momentum fraction $x$.  We find
  that the predicted fluxes vary up to almost two orders of magnitude at the
  largest energies studied, depending on the chosen extrapolation of the PDF's.
  We show that the spectral index of the atmospheric leptonic fluxes depends
  linearly on the slope of the gluon distribution function at very small $x$.
  This suggests the possibility of obtaining some bounds on this slope in
  ``neutrino telescopes'', at values of $x$ not reachable at colliders,
  provided the spectral index of atmospheric leptonic fluxes could be
  determined.
\end{abstract}
 
 
\section{Introduction} 
 
\label{sect:intro} 
 
The flux of atmospheric neutrinos and muons at very high energies, above $
1~{\rm TeV}$, originates primarily from semileptonic decays of charmed
particles instead of pions and kaons, which are the dominant decay modes at
lower energies (see for example \cite{g1}). This flux is one of the most
important backgrounds for ``neutrino telescopes'', limiting their sensitivity
to astrophysical signals, especially for future ${\rm {km}^{3}}$ detectors
which might be able to observe neutrinos and muons at extremely high energies,
even up to $10^{12}$ GeV.
 
We use perturbative QCD (pQCD), the theoretically preferred model, to compute
the charm production. We perform a true next-to-leading order (NLO) pQCD
analysis of the production of charmed particles in the atmosphere, together
with a full simulation of the particle cascades down to the final muons and
neutrinos. This is done by combining the NLO pQCD calculations of charm
production and computer routines of Mangano, Nason, and Ridolfi \cite {nde,MNR}
(called MNR in the following) with the computer simulations of the cascades
generated by PYTHIA \cite{PYTHIA}. These are the same programs currently used
to compare pQCD predictions with experimental data in accelerator experiments.
 
We have already presented results of our calculations in a previous paper
\cite{GGV1} (called GGV1 from now on), in which all the details of the program
we use can be found. The main goal of our first paper was to compare the fluxes
obtained with the NLO and the leading order (LO) calculations, i.e. we computed
the $K$ factor for the neutrino and muon fluxes. This was done to improve on
the first study of atmospheric fluxes based on pQCD, performed by Thunman,
Ingelman, and Gondolo a few years ago in Ref.~\cite{TIG} (called TIG in the
following). TIG used the LO charm production cross section computed by PYTHIA,
multiplied by a constant $K$ factor of 2 to bring it in line with the NLO
values, and supplemented by parton shower evolution and hadronization according
to the Lund model.
 
In GGV1 we found the $K$ factors for different parton distribution functions
(PDF's), as function of energy, to be in a range between 2.1 and 2.5.  A similar
analysis was recently made in Pasquali, Reno, and Sarcevic \cite{PRS} (called
PRS from now on), with results compatible with ours, using a treatment of the
problem complementary to ours.  In fact, PRS used approximate analytic
solutions to the cascade equations in the atmosphere, also introduced by TIG,
while we make instead a full simulation of the cascades.
 
In GGV1 we showed that the approach used by TIG (i.e. multiplying the LO fluxes
by an overall $K$ factor of 2) was essentially correct, except for their
relative low $K$ factor (since $K$ values of 2.2 - 2.4, depending slightly on
the PDF, provide estimates of the NLO within about 10\%). However, while TIG
found neutrino and muon fluxes lower than the lowest previous estimate, we
found instead larger fluxes (by factors of 3 to 10 at the highest energies,
about $10^9$ GeV), in the bulk part of previous predictions. The main reason
for this difference is studied in this paper.
 
Here we explore the dependence of the atmospheric fluxes on the extrapolation
of the gluon PDF at very small partonic momentum fraction $x$, $x \lesssim
10^{-5}$, which is crucial for the fluxes at high energies. As explained below,
the relevant momentum fraction $x$ of the interacting atmospheric parton is of
the order of the inverse of the leptonic energy $E_l $ (in the atmospheric rest
frame) in GeV. This energy, in turn, is of the order of 0.1 $E$, where $E$ is
the energy per nucleon of the incoming cosmic ray in the lab.\ frame (the
atmospheric rest frame). Thus, for $E_l \gtrsim 10^5 $ GeV, we need the PDF's at
$x \lesssim 10^{-5}$, values of $x$ which are not reached experimentally. The
final fluxes depend mostly on the gluon PDF, since this is by far the dominant
one at these small $x$ values and charm is mostly produced through gluon-gluon
fusion processes.
 
A concern that has been expressed to us several times is the applicability of
the MNR NLO-pQCD calculations, mostly done for accelerator physics, to the
different kinematic domain of cosmic rays. In response we remark that, for the
less steep extrapolations of the gluon structure function $g(x)$ that we use at
small $x$, we have large logarithms, known as ``ln(1/x)" terms, where $x\simeq
\sqrt{4 m_c^2/ s}$, $s$ is the hadronic center of mass squared energy and this
$x$ is the average value of the hadron energy fraction needed to produce the $c
\bar{c }$ pair. With the extrapolation $ g(x) \simeq x^{\lambda -1}$ (see
below) and $\lambda$ close to 0.5, and possibly for the intermediate choices of
$\lambda$ also, there should be no large logarithm. The problem arises for
$\lambda$ too close to zero. We will attempt to deal with this problem in
future work. Moreover, contrary to the case in accelerators, we do not have the
uncertainty present in the differential cross sections \cite {MNR} when $k_T$
is much larger than $m_c$, due to the presence of large logarithms of $(k^2_T +
m_c^2)/m_c^2$. Because we do not have here a forward cut in acceptance, the
characteristic transverse charm momentum in our simulations is of the order of
the charm mass, $k_T \simeq O(m_c)$.
 
In this paper, as in GGV1, the MNR program is used to compute the inclusive
charm cross section and the cascades simulated by PYTHIA are initiated by a
single $c$ quark. This is the `single' mode described in our previous paper
GGV1, where we argued its advantages. We explained there our normalization of
the NLO charm production cross sections in the MNR program, and described in
detail the computer simulations used to calculate the neutrino and muon fluxes,
which we briefly review in Sections~\ref{sect:NLO} and \ref{sect:simulation}. Except for the inclusion of the
NLO calculations our model closely follows TIG. In Section~\ref{sect:fluxes} we show the
neutrino and muon fluxes we obtain for different low $x$ behaviors of the gluon
PDF and we compare them with the TIG fluxes. In Section~\ref{sect:analytic}, we
give analytic arguments that explain and support our results.

Finally, as in GGV1 (and TIG), we consider only vertical showers for
simplicity.  We intend to study those from all directions in the future.
 
\section{Charm production in pQCD and choice of PDF's} 
 
\label{sect:NLO} 
 
Our NLO calculation is based on the MNR computer code. The NLO cross section
for charm production depends on the choice of the parton distribution functions
and on three parameters: the charm quark mass $m_c$, the renormalization scale
$\mu_R$, and the factorization scale $\mu_F$. In order to calibrate the charm
production routines we fit the most recent experimental data
\cite{alves1,alves2,ada,fmnr} (differential and total cross sections) with one
and the same combination of $m_c$, $\mu_R$, and $ \mu_F$, for each PDF we use
(see \cite{GGV1} for complete details). Several choices of $m_c$, $\mu_R$ and
$\mu_F$ may work equally well. In fact the cross sections increase by
decreasing $\mu_F$, $\mu_R$ or $m_c$, so changes in the three variables can be
played against each other to obtain practically the same results. We use just
one such choice for each PDF. We intend to further study the uncertainty
related to this range of possible choices in the future.
 
As in GGV1, here we use the PDF's MRS R1, R2 \cite{MRS1} and CTEQ 4M \cite
{CTEQ}, with the following parameters. We choose $\mu_{R} = m_T$, $\mu_{F} = 2
m_T$ for all sets, where $m_T$ is the transverse mass, $m_{T} = \sqrt{
  k_{T}^{2} + m_{c}^{2}}$, and
\begin{eqnarray} 
m_{c}&=&1.185~{\rm GeV} \quad\hbox{for MRS R1,}  \label{eq:1} \\ 
m_{c}&=&1.31~{\rm GeV} \quad\hbox{for MRS R2,}  \label{eq:1b} \\ 
m_{c}&=&1.27~{\rm GeV} \quad\hbox{for CTEQ 4M.}  \label{eq:1c} 
\end{eqnarray} 
 
The data we use for this `calibration' of the MNR program are shown in Table 1
and Table 2 of GGV1. In this paper, we add to our list of PDF's the latest of
the MRS set, the MRST \cite{MRST}, with charm mass
\begin{equation} 
m_{c}~=~1.25~{\rm GeV} \quad\hbox{for MRST,}  \label{eq:1d} 
\end{equation} 
obtained with the same procedure used for the other PDF's. 
 
As we will see clearly in Sect.~\ref{sect:analytic}, due to the steep decrease
with increasing energy of the incoming flux of cosmic rays, only the most
energetic charm quarks produced count, and these come from the interactions of
projectile partons carrying a large fraction of the incoming nucleon momentum.
Thus, the characteristic $x$ of the projectile parton, that we call $x_1$, is
large. It is $x_1 \simeq O(10^{-1})$. We can, then, immediately understand that
very small partonic momentum fractions are needed in our calculation, because
typical partonic center of mass energies $ \sqrt{\hat s}$ are close to the $c
\bar c$ threshold, $2m_c \simeq 2$ GeV (since the differential cross section
decreases with increasing ${\hat s}$) while the total center of mass energy
squared is $s = 2 m_N E$ (with $m_N$ the nucleon mass, $m_N \simeq 1$ GeV).
Calling $x_2$ the momentum fraction of the target parton (in a nucleus of the
atmosphere), then $x_1 x_2 \equiv \hat s/s = 4 m_c^2/(2 m_N E) \simeq$ GeV/$E$.
Thus, $x_2 \simeq O({\rm GeV}/0.1E $), where $E$ is the energy per nucleon of
the incoming cosmic ray in the lab.\ frame. The characteristic energy $E_c$ of
the charm quark and the dominant leptonic energy $E_l$ in the fluxes are $E_l
\simeq E_c \simeq 0.1 E $, thus $x_2 \simeq O({\rm GeV}/E_l$). Namely $x_2
\simeq 10^{-6}, 10^{-7}$ at $E_l \simeq 1, 10~{\rm PeV}$.
 
For $x > 10^{-5}$ ($E \lesssim 10^3 $ TeV), PDF's are available from global
analyses of existing data. We use four sets of PDF's. Three of these, MRS R1,
MRS R2 \cite{MRS1} and CTEQ 4M \cite{CTEQ} (used also in GGV1), incorporate
most of the latest HERA data and cover the range of parton momentum fractions
$x \geq 10^{-5}$ and momentum transfers $Q^2 \geq 1.25-2.56 {\rm~GeV}^2$.
MRS R1 and MRS R2 differ only in the value of the strong coupling constant
$\alpha_s$ at the Z boson mass: in MRS R1 $\alpha_s (M_Z^2) = 0.113$ , and in
MRS R2 $\alpha_s (M_Z^2) = 0.120$. The former value is suggested by ``deep
inelastic scattering'' experiments, and the latter by LEP measurements. This
difference leads to different values of the PDF parameters at the reference
momentum $Q_0^2 = 1.25~{\rm GeV}^2$, where the QCD evolution of the MRS R1 and
R2 PDF's is started. The CTEQ 4M is the standard choice in the $\overline{MS} $
scheme in the most recent group of PDF's from the CTEQ group ($\alpha_s (M_Z^2)
= 0.116$ for CTEQ 4M). In this paper we also use the very recent MRST
\cite{MRST}. This new PDF set includes all the latest experimental measurements
that have become available and, for the first time, an investigation of the
uncertainty in the gluon distribution function. We will use the main choice of
the MRST set, the ``central gluon" MRST, the central value of the gluon PDF's of
the package, which is considered the optimum global choice of this new set. The
range in $ Q^2$ and $x$ of MRST set is the same as for the older MRS R1-R2 ($x
\geq 10^{-5}$ and $Q^2 \geq 1.25~{\rm GeV}^2$), and $\alpha_s (M_Z^2) =
0.1175$.
 
For $x\ll 1$, all these PDF's go as  
\begin{equation} 
x f_i (x, Q^2) \simeq A_i x^{-\lambda_i(Q^2)},  \label{eq:2} 
\end{equation} 
where $i$ denotes valence quarks $u_v, d_v$, sea quarks $S$, or gluons $g$.
The PDF's we used have $\lambda_S (Q_0^2) \not = \lambda_g (Q_0^2)$, in contrast
to older sets of PDF's which assumed an equality. As $x$ decreases the density
of gluons grows rapidly. At $x \simeq 0.3$ it is comparable to the quark
densities but, as $x$ decreases it increasingly dominates over them. Quark
densities become negligible at $x \lesssim 10^{-3}$.
 
The PDF's need to be extrapolated to $x < 10^{-5}$ ($E \gtrsim 10^3 $ TeV).
Extrapolations based on Regge analysis usually propose $x g(x) \sim
x^{-\lambda}$ with $\lambda \simeq 0.08$ \cite{lowx}, while evolution equations
used to resum the large logarithms $\alpha_s \ln (1/x)$ mentioned before, such
as the BFKL (Balitsky, Fadin, Kuraev, Lipatov \cite{BFKL}) find also $xg(x)
\sim x^{-\lambda}$, but with $\lambda \simeq 0.5$.
 
In this work we use extrapolations with different values of $\lambda$. For the
older MRS R1-R2 and CTEQ 4M we consider only the two extreme behaviors and the
intermediate one that we used in GGV1, namely: (i) a constant extrapolation
$\lambda_g(Q^2) = 0$ for $x \leq10^{-5}$; (ii) a linear extrapolation of $\ln
g(x)$ as a function of $\ln x$, $\ln g(x) = -( \lambda_g(Q^2) + 1) \ln x + \ln
A_g $, where $\lambda_g(Q^2)$ is taken at $x = 10^{-5}$, the smallest $x$ for
which the PDF's are provided (we call $\lambda$(R1), $\lambda$(R2) or
$\lambda$(4M) the $\lambda$'s so obtained); (iii) an extrapolation with
$\lambda_g(Q^2) = 0.5$ for $x \leq 10^{-5}.$ Cases (i) and (iii) are extreme
choices theoretically justified before \cite{lowx}, while (ii) is somewhat in
between, with a resulting $\lambda \simeq 0.2-0.3$.
 
For the new MRST we have included several values of $\lambda$, in order to test
the dependence on this parameter in a more complete way: (i) extrapolations
with different $\lambda$'s, i.e. $\lambda_g(Q^2) = 0, 0.1, 0.2, 0.3, 0.4, 0.5$
for $x \leq10^{-5}$; (ii) we also included the linear extrapolation of $\ln
g(x)$ as a function of $\ln x$, similar to the second intermediate choice of
the previous list; we will call $\lambda$(T) the $\lambda$ obtained in this
way.
 
\section{Simulation of particle cascades in the atmosphere} 
 
\label{sect:simulation} 
 
In this section we briefly describe the computer simulation used to calculate
the neutrinos and muons fluxes; a more detailed description can be found in
GGV1 \cite{GGV1}. The charm production process in the atmosphere and the
particle cascades are simulated by modifying and combining together two
different programs: the MNR routines \cite{MNR} and PYTHIA 6.115 \cite{PYTHIA}.
 
The MNR program was modified to become an event generator for charm production
at different heights in the atmosphere and for different energies of the
incoming primary cosmic rays.
 
The charm quarks (and antiquarks) generated by this first stage of the program
are then fed into a second part which handles quark showering, fragmentation
and the interactions and decays of the particles down to the final leptons. The
cascade evolution is therefore followed throughout the atmosphere: the muon and
neutrino fluxes at sea level are the final output of the process.
 
In order to make our results comparable to those of TIG, we keep the same
modeling of the atmosphere and of the primary cosmic ray flux as in TIG and the
same treatment of particle interactions and decays in the cascade.
 
We recall however that our main improvements are the inclusion of a true NLO
contribution for charm production, the use of updated PDF's and, in this second
paper, the different extrapolations used for the gluon PDF at low $x$.
 
In the rest of this section we review briefly the model for the atmosphere and
the primary flux used in this study, which is the same of GGV1 and was
introduced originally by TIG.
 
We assume a simple isothermal model for the atmosphere. Its density at vertical
height $h$ is
\begin{equation} 
\rho (h) = {\frac{X_0 }{h_0}} \, e^{-h/h_0},  \label{eq:3} 
\end{equation} 
with the parameters, scale height $h_0 = 6.4~{\rm km} $ and column density $
X_0 = 1300~{\rm g/cm^2} $ at $h=0$, chosen as in TIG to fit the actual density
in the range $3~{\rm km} < h < 40~{\rm km}$, important for cosmic ray
interactions. Along the vertical direction, the amount of atmosphere traversed
by a particle, the depth $X$, is related to the height $h$
simply by 
\begin{equation} 
X = \int_h^\infty \rho(h^\prime) dh^\prime = X_0 e^{-h/h_0}.  \label{eq:4} 
\end{equation} 
The atmospheric composition at the important heights is approximately constant:
78.4\% nitrogen, 21.1\% oxygen and 0.5\% argon with average atomic number
$\langle A \rangle$ = 14.5.
 
Following TIG \cite{TIG}, we neglect the detailed cosmic ray composition and
consider all primaries to be nucleons with energy spectrum
\begin{eqnarray}
  \phi_N (E,0) \left
 [\rm{nucleons}\over {cm^2~ s ~sr ~GeV~/A} \right ]
  =\phi_0 E^{-\gamma-1}= \\&& \hskip-8em
= \begin{cases}
  {1.7 (E/{\rm GeV})^{-2.7} & for $ E < 5~10^6$ GeV\cr 
  174 (E/{\rm GeV})^{-3} & for $ E > 5~10^6$ GeV
    \cr}
\end{cases}\nonumber
\end{eqnarray}

The primary flux is attenuated as it penetrates into the atmosphere by
collisions against the air nuclei. An approximate expression for the intensity
of the primary flux at a depth $X$ is (see \cite{TIG} again)
\begin{equation} 
\phi_N (E, X) = e^{- X/\Lambda_N}~ \phi_N(E, 0)~.  \label{eq:6} 
\end{equation} 
The nuclear attenuation length $\Lambda_N$, defined as  
\begin{equation} 
\Lambda_N(E) = {\frac{\lambda_N(E) }{1 - Z_{NN} (E)}} ~,  \label{eq:7} 
\end{equation} 
has a mild energy dependence through $\lambda_N$ and $Z_{NN}$, the
spectrum-weighted moment for nucleon regeneration in nucleon-nucleon
collisions. We use the $Z_{NN}$ values in Fig.~4 of Ref. \cite{TIG}. The
interaction thickness $\lambda_N$ is
\begin{equation} 
\lambda_N(E,h) = {\frac{\rho(h)}{\sum_{A} \sigma_{NA}(E) n_A(h)}}~, 
\label{eq:8} 
\end{equation} 
where $n_A(h)$ is the number density of air nuclei of atomic weight $A$ at
height $h$ and $\sigma_{NA}(E)$ is the total inelastic cross section for
collisions of a nucleon $N$ with a nucleus $A$.  This cross section scales
essentially as $A^{2/3}$, $\sigma_{NA}(E) = A^{2/3} \sigma_{NN} (E)$. For $
\sigma_{NN}(E)$ we use the fit to the available data in Ref. \cite{RPP96}.
Using our height independent atmospheric composition, we simplify Eq.~(\ref
{eq:8}) as follows,
\begin{equation} 
\lambda_N(E,h) = {\frac{\langle A \rangle }{\langle A^{2/3} \rangle}} \, {
\frac{{\rm u} }{\sigma_{NN} (E) }} = 2.44 \, {\frac{ {\rm u} }{\sigma_{NN}(E)
}} ~.  \label{eq:9} 
\end{equation} 
Here $\langle ~ \rangle$ denotes average and u is the atomic mass unit, that we
write as
\begin{equation} 
{\rm u} = 1660.54 {\rm ~mb~g/cm^2}.  \label{eq:10} 
\end{equation} 
Therefore in our approximations $\lambda_N(E)$ is independent of height.
 
\section{Neutrino and muon fluxes} 
 
\label{sect:fluxes} 
 
We present here the results of our simulations with all the PDF's and the values
of $\lambda$ described in Section~\ref{sect:NLO}.
 
The NLO total inclusive charm-anticharm production cross sections
$\sigma_{c\bar c}$ for our four different PDF's are shown in Fig.~1 over the
energy range needed by our program, $E \leq 10^{11}$ GeV. In the top part of
the figure we compare the results of MRS R1-R2 and CTEQ 4M (with their
different values of $\lambda$ described before) to the cross section used in
the TIG model.  In the bottom part we show the same comparison, done just with
the new MRST, with its different $\lambda$'s (in all these figures cross
sections increase for increasing values of $\lambda$).
 
All these cross sections were calculated using the MNR program, with the
`calibration' described in Sect.~\ref{sect:NLO}, up to the NLO contribution.
We can see in the figure that all our cross sections agree at low energies, as
expected due to our `calibration' at 250 GeV, and are very similar for energies
up to $10^{6}$ GeV. Beyond this energy they start showing their dependence on
the $\lambda$ value and also a slight dependence on the PDF used, which was
already noticed in GGV1. As it can be seen from both parts of the figure, the
increase of the cross sections with $\lambda$ is evident at the highest
energies: at the maximum energy considered the cross sections for the two
extreme values of $\lambda$ differ by almost a factor of ten.
 
We also notice that, for energies above $10^{4}$ GeV, our cross sections are
always considerably higher than the one used by TIG. As we have already
explained in GGV1, TIG used an option of PYTHIA by which the gluon PDF is
extrapolated for $x \leq 10^{-4}$ with $\lambda=0.08$. In fact the TIG cross
section at the highest energies shows the same slope of our results for $
\lambda \simeq 0$, but is always lower than our lowest cross sections by about
a factor of three.
 
This can be explained only in part by the fact that the TIG cross section up 
to NLO is the LO result obtained with PYTHIA, multiplied by a constant $K$ 
factor of 2, while at large energies the $K$ factor (see GGV1 for details) 
is actually larger than 2 by about 10-15\%. The bulk of the difference is 
however due to the different evaluations of the cross sections, even at LO, 
done by the MNR routines (our method) and directly by PYTHIA (approach used 
by TIG). 
 
Our results for the prompt fluxes are shown in Figs.~2--5, for MRS R1-R2, CTEQ
4M and MRST.
 
In Figs.~2 and~3 we show the $E_{\ell}^{3}$-weighted vertical prompt fluxes
$E_{\ell}^3 \phi_\ell$, calculated to NLO, for muons and muon-neutrinos,
together with the fluxes from TIG, both from prompt and conventional sources
(dotted lines). The flux of electron-neutrinos is practically the same as that
of muon-neutrinos.  Fig.~4 describes the spectral index of the differential
fluxes, defined as $\alpha_{\ell} = -\partial \ln \phi_{\ell}/\partial \ln
E_{\ell}$.

The effects of the different extrapolations of $g(x)$ to $x<10^{-5}$ (see
Sect.~\ref{sect:NLO}) are noticeable at $E_{\ell} \gtrsim 10^{5}~{\rm GeV}$. In Figs.~2
and~3, the $E_{\ell}^{3}$-weighted fluxes increase with $\lambda $: they can
differ by up to two orders of magnitude at the highest energy considered,
$10^{9}$ GeV, for the two extreme choices of $\lambda $. This behavior is
similar for all the PDF's considered.
 
The $\lambda$ dependence of the fluxes can also affect the energy at which the
prompt contribution dominates over the conventional sources: this is
particularly true for the muon fluxes as it can be seen in Fig.~2; for the
$\nu_\mu + \bar {\nu}_\mu$ fluxes this effect is less important (see Fig.~3)
and it doesn't exist for the $\nu_e+ \bar {\nu_e}$ fluxes, for which the
conventional contribution is much lower. Apart from these differences due to
the $\lambda$ values, charm decay dominates over conventional sources at $E_\mu
\gtrsim 10^6~{\rm GeV}$ for muons, $E_{\nu_\mu} \gtrsim 10^5~{\rm GeV}$ for
muon-neutrinos, and $E_{\nu_e} \gtrsim 10^4~{\rm GeV}$ for electron-neutrinos.
 
We also see that all our fluxes for $\lambda \simeq 0$ are similar to those of
TIG at energies above $10^6$ GeV. We have already mentioned that TIG used a
very low value of $\lambda$, $\lambda=0.08$. It is remarkable that, for these
low values of $\lambda$, we obtain similar final fluxes in spite of the
differences of the two simulations and of the total cross sections already
noted in Fig.~1.
 
We can also compare our fluxes to those of the recent PRS results \cite{PRS}.
As we have already noticed in GGV1, for intermediate values of $\lambda$ our
results are very similar to the PRS ones. From Fig.~3, for example, we see
that our fluxes for the $\lambda = 0.3$ case (calculated with MRST) are close
to the corresponding PRS results shown in Fig.~8 of Ref. \cite{PRS}, calculated
with CTEQ 3M and $\lambda \simeq 0.3$. Our results are lower than the PRS by
$30-50 \%$ at the highest energies, which is probably due to the PDF's used and
to the different approach of the two groups.
 
Regarding the dependence of the spectral index $\alpha_{\ell}$ on the slope
$\lambda$ of the gluon PDF, we notice in Fig.~4 that, for all four PDF's, above
about $10^6$ GeV the differences in slope between the $\lambda = 0$ and $
\lambda = 0.5$ fluxes is about 0.5, suggesting that the spectral index is $
\alpha_{\ell}(E_{\ell}) = b_{\ell}(E_{\ell}) - \lambda$, namely, 
\begin{equation} 
\phi_{\ell}(E_{\ell}) \sim E_{\ell}^{~- \alpha_{\ell}(E_{\ell})}~=~ 
E_{\ell}^{~- b_{\ell}(E_{\ell}) + \lambda}~,  \label{eq:11} 
\end{equation} 
where $b_{\ell}(E_{\ell})$ is an energy dependent coefficient, that can be read
off directly from the $\lambda = 0$ curve ($b_{\ell}(E_{\ell})$ is the spectral
index for $\lambda = 0$). We will justify this result in Sect.~\ref{sect:analytic}. Due to this
linear dependence of the spectral index on $\lambda$, given a model which
specifies the function $b_{\ell}(E_{\ell})$, the value of $\lambda$ could be
determined through a measurement of any of the $\phi_{\ell}$ fluxes at two
different energies. We will study in detail this possibility elsewhere
\cite{GGV3}.
 
Here we only comment on the typical rates in a km$^3$ detector. It can be estimated from the curves of Fig.~2 that the number of prompt atmospheric muons traversing a km$^3$ detector from above
would be over 100 per year around a muon energy of 1 PeV, decreasing rapidly to
less than 1 per year above 100 PeV. In this energy range there is a concrete
possibility of detecting these prompt muons. Notice that the intensity of the
prompt muon flux depends critically on the value of $\lambda$, suggesting still
another way to estimate $\lambda$ through the measurement of the fluxes.
 
In Fig.~5 we study the dependence of the prompt fluxes on the PDF for fixed
values of $\lambda$. We summarize our previous results for $\lambda=0$ (left)
and for $\lambda=0.5$ (right), and compare them again to TIG. The figures on
the top show the $E_{\ell}^{3}$-weighted fluxes, those on the bottom the spectral indices.
As we already noticed in GGV1, the dependence on the PDF is not strong, all
fluxes are very similar. This indicates that our procedure for the
`calibration' of our simulation with different PDF's (described in Sect.~\ref{sect:NLO}) is
good. There are, however, some differences between the PDF's: in some cases
(especially for $\lambda=0$) the results of MRS R2 and
CTEQ 4M are very similar and higher than those of MRS R1 and MRST (also very
close to each other). The maximum difference between all these fluxes is at the
level of 30 to 70\% at high energies.
 
We want here to remark once more that our $\lambda=0$ fluxes are very close to
that of TIG at energies above $10^6$ GeV (and also below $10^3$ GeV, but the
prompt fluxes are not important at these low energies). For increasing values
of $\lambda$, our results are higher than TIG, even by two orders of magnitude
for $\lambda=0.5$ and at the highest energies. From the bottom part of the figure we notice that also the spectral indices are almost independent of the PDF used. This indicates that the linear dependence between $\alpha_{\ell}$ and $\lambda$ of Eq.~(\ref{eq:11}) is not affected by the choice of the PDF and again might be used to determine the value of $\lambda$. We will return on this analysis in more details in another paper \cite{GGV3}.
 
\section{Analytic insight} 
 
\label{sect:analytic} 
 
In this section we first find the characteristic values of the partonic
momentum fractions in the cosmic ray nucleus and in the nucleus in the
atmosphere, and then derive the linear relation between the slope of the
atmospheric muon (or neutrino) fluxes and the slope of the gluon parton
distribution function. 

We first show that the characteristic values of the partonic momentum
fractions of the incoming cosmic ray parton, $x_1$, and of the target parton
belonging to a nucleus in the atmosphere, $x_2$, are respectively, 
\begin{equation} 
x_1 \simeq 10^{-1} \ \ \ x_2 \simeq {\ (E/10~{\rm GeV})^{-1}}  \label{eq:12} 
\end{equation} 
where $E$ is the energy of the incoming nucleon (a proton in this paper) in the
atmosphere reference frame. Precisely because of the small value of $x_2$, for
the relevant energies $E \gtrsim 10^4~{\rm GeV}$ the gluon density $g (x_2)$
is much larger than the density of quarks, which we, thus, neglect in these
analytic arguments. 
 
Let us first consider the charm flux at production $d\phi_c (E_c, X)/dX$,
defined as the rate of c quark production\footnote{ This is what we compute in
  our simulations (we use our `single' mode), only the production of a c quark
  is calculated. Then the result is multiplied by two to include the
  contribution of the antiquark (see \cite{GGV1} for details).} per unit area,
unit depth and unit charm energy ($E_c$ in the atmosphere reference frame) in
the interactions of the attenuated nucleon flux $\phi_N(E, X)$ with the air
nuclei in the atmospheric layer between $X$ and $X + dX$. To obtain $d\phi_c
(E_c, X)/dX$ for a layer of transverse area ${\cal A}$ and height $|dh|$, we
simply multiply the $c$ production rate per air nucleus (which equals the
incoming nucleon flux at depth $X$ times the cross section for $N+A \to c + Y$,
where $Y$ stands for ``anything" and N is simply a proton p in our study) by
the number of nuclei $A$ in the layer (which is ${\cal A} |dh| n_A(h)$) and
divide the result by the transverse area ${\cal A}$ and the layer thickness $dX
= \rho(h) |dh|$. We find
\begin{equation} 
{\frac{d\phi_c(E_{{\rm c}},X)}{dX}} = \sum_{{\rm A}} {\frac{n_{{\rm A}}(h) }{
\rho(h)}} \int_{E_{{\rm c}}}^\infty dE \, \phi_{{\rm N}}(E,X) \, {\frac{
d\sigma({\rm pA\to cY;E,E_c}) }{dE_{{\rm c}}}} .  \label{eq:13} 
\end{equation} 
 
We assume that the charm production cross section simply scales as $A$, which
is expected when it is much smaller than the total inelastic cross section. In
this case, the sum over $A$ becomes trivial, and we have (u is the atomic mass
unit)
\begin{eqnarray} 
{\frac{d\phi_c(E_{{\rm c}},X)}{dX}} & = & {\frac{1}{{\rm u}}} 
\int_{E_c}^{\infty} dE ~\phi_N(E,X) \, {\frac{d\sigma (pN \rightarrow 
cY;E,E_c) }{dE_c}}~.  \label{eq:14} 
\end{eqnarray} 
 
In these analytical considerations, we assume a simple power law for the
primary flux and an energy independent attenuation length.\footnote{ The
  dependence of $\Lambda_N$ on $E$ is actually very mild. In fact the whole
  factor $e^{- X/\Lambda_{N}(E)}$ behaves like $E^{-\beta}$ with $\beta \simeq
  0.1$ for $E\gtrsim 10^6~{\rm GeV}$ and $\beta$ even smaller for $E\lesssim
  10^6~{\rm GeV}$. Including this contribution in our analytic argument would
  just mean to replace $\gamma$ with $\gamma+\beta$ everywhere, i.e. the total
  spectral index would become $\gamma+1+\beta\simeq 3.1$ instead of $3.0$, for
  energies above the knee at $E=5~10^6~{\rm GeV}$. This slight change can 
  actually be seen in our results of Fig.~7b (see the description of that
  figure).} With these approximations, the attenuated primary flux reads (see
Eqs.~8-13)
\begin{equation} 
\phi_N(E,X) = \phi(X) E^{-\gamma-1},  \label{eq:15} 
\end{equation} 
where $\phi(X) = \phi_0 \exp(-X/\Lambda_N)$. Substituting this approximate
expression for the attenuated primary flux and changing the integration
variable from $E$ to $x_E = E_c/E$ in Eq.~(\ref{eq:14}), we find
\begin{equation} 
{\frac{d\phi_c(E_{{\rm c}},X)}{dX}} = {\frac{\phi(X) }{{\rm u}}} \, 
E_c^{-\gamma-1} \int_0^1 dx_E \, x_E^\gamma \, {\frac{d\sigma (pN \to 
cY;x_E,E_c) }{dx_E}}~.  \label{eq:16} 
\end{equation} 
 
The differential cross section $d \sigma(pN \to cY)/dx_E$ is given in terms of
the partonic differential cross section $d {\hat{\sigma}}_{ij}/dx_E$ (where $i$
and $j$ are partons belonging to the projectile 1 and the target 2
respectively), and the PDF's $f^1_i (x_1,\mu_F^2)$ and $f^2_j (x_2,\mu_F^2)$
as 
\begin{equation} 
\frac{d \sigma (pN \to cY)}{dx_E} = \sum_{ij} \int dx_1 dx_2 f^1_i 
(x_1,\mu_F^2) f^2_j(x_2,\mu_F^2) \frac{d {\hat{\sigma}}_{ij}}{dx_E}. 
\label{eq:17} 
\end{equation} 
 
Here $x_1$ and $x_2$ are the momentum fractions of the projectile and target
partons. Mangano et al. \cite{MNR} give the partonic cross section in terms
of functions $h_{ij}$ as 
\begin{equation} 
E_c \frac {d {\hat{\sigma}}_{ij}} {d^3 k} = \frac{\alpha^2_s(\mu_R)}{\hat{s}
^2}~ h_{ij}(\tau_x, \tau_2, \rho, \mu_R, \mu_F),  \label{eq:18} 
\end{equation} 
where $k$ and $E_c$ are the momentum and energy of the produced $c$ quark, and,
in the notation of Ref.~\cite{MNR}, $\rho \equiv 4 m^2_c/{\hat{s}}$, $ \tau_x =
1 - \tau_1 - \tau_2$, $\tau_1 \equiv (k \cdot p_1/p_1 \cdot p_2)$, $ \tau_2
\equiv ( k \cdot p_2/p_1 \cdot p_2)$ and $\hat{s} \equiv (p_1+p_2)^2$ , while
$p_1$ and $p_2$ are the projectile and target parton momenta respectively,
$p_1=x_1 P_1, p_2=x_2 P_2$. The hats indicate quantities in the partonic center
of mass (those without hats are in the lab.\ frame at rest with the
atmosphere).
 
In the partonic center of mass frame, the projectile and target parton momenta
are
\begin{equation} 
\hat{p_1}= \left ( {\frac{\sqrt{\hat{s}} }{2}}, 0,0, {\frac{\sqrt{\hat{s}} }{
2}}\right ),~ ~\hat{p_2}= \left( {\frac{\sqrt{\hat{s}} }{2}}, 0,0, -{\frac{
\sqrt{\hat{s}} }{2}} \right ),~~ \hat{k}= \left( {\hat {E}}_c, 0,{\hat{k}}
_T, {\hat{k}}_{\/\/} \right ), 
\end{equation} 
and we have  
\begin{equation} 
\tau_2 = {\frac{ {\hat{E}}_c + {\hat{k}}_{\/\/} }{\sqrt{\hat{s}} }},~~ 
\tau_x = 1 -{\frac{ 2{\hat{E}}_c }{\sqrt{\hat{s}} }}. 
\end{equation} 
Then, after integration over azimuthal angles, 
\begin{equation} 
{\frac{ d^3 k }{E_c}} = {\frac{ d^3 {\hat {k} } }{{\ \hat {E_c}} }} = 2 \pi 
d {\ \hat {E_c} } d{\hat{k}}_{\/\/} = \pi {\hat{s}} d\tau_2 d\tau_x . 
\label{eq:19} 
\end{equation} 
 
The kinematic bounds $m_c \leq {\hat{E}}_c \leq \sqrt{\hat{s}}/2 $ and $|{
  \hat{k}}_{\/\/} | \leq \sqrt{{\hat{E}}_c^2 - m_c^2}$ fix the integration
domains of $\tau_2$ and $\tau_x$. Using $\rho = 4 m_c^2 / {\hat{s}}$, we get $(
1- \sqrt{ 1- \rho})/2 \leq \tau_2 \leq ( 1 + \sqrt{ 1- \rho})/2$ and $0 \leq
\tau_x \leq 1 - \tau_2 - (\rho / 4 \tau_2)$.  We can use the relation
\begin{equation} 
x_E = {\frac{E_c }{E}} = {\frac{k \cdot P_2 }{P_1 \cdot P_2}} = x_1 {\frac{k 
\cdot p_2 }{p_1 \cdot p_2}} = x_1 \tau_2,  \label{eq:20} 
\end{equation} 
to write the differential cross section in $dx_E$ as 
\begin{equation} 
\frac{d {\hat{\sigma}}_{ij}}{dx_E} = \int d^3 k \, \frac{d {\hat{\sigma}}
_{ij}}{d^3 k} \, \delta (x_E - x_1 \tau_2).  \label{eq:21} 
\end{equation} 
 
The bound $x_1 x_2 = {\hat{s}}/ s \geq 4 m_c^2/2 m_p E = 4 \epsilon x_E$ ($m_p$
is the proton mass, $m_p \simeq 1~{\rm GeV}$), where we define
\begin{equation} 
\epsilon = {\frac{m_c^2 }{2m_p E_c}},  \label{eq:23} 
\end{equation} 
implies that $x_1$ and $x_2$ have a minimum lower bound larger than zero. In
fact, $x_1 \geq 4 \epsilon x_E / x_2 \geq 4 \epsilon x_E$ (since $x_2 \leq 1$).
Taking $x_1$ as the independent variable, then $4 \epsilon x_E \leq x_1 \leq 1$
and $4 \epsilon x_E/x_1 \leq x_2 \leq 1$. We now change the order of the
integrations, in order to perform the integration in $x_E$ before the
integrations in $x_1$, $x_2$ and $\tau_2$.
 
The integration over $x_E$ in Eq.~(\ref{eq:16}) then becomes trivial, amounting
to the replacement of $x_E^\gamma$ by $x_1^\gamma \tau_2^\gamma$, except for
the necessary changes in the integration domains which become $0 \leq
x_1,x_2,\tau_2 \leq 1$ and $0 \leq x_E \leq ( x_1 x_2 / \epsilon)
\tau_2(1-\tau_2)$. For the $\delta(x_E-x_1\tau_2)$ in Eq.~(\ref{eq:21}) to yield
a non-zero result, we need to take $0 \leq x_1 \tau_2 \leq ( x_1 x_2 /
\epsilon) \tau_2(1-\tau_2)$, which means that $\tau_2 \leq 1- (\epsilon/x_2)$ ,
and given that $\tau_2 \geq 0$, this means $x_2 \geq \epsilon$.  This leads to
a factorization of the $x_1$ and $x_2$ integrations as follows:
\begin{eqnarray} 
\int^1_0 dx_E ~x^{\gamma}_E~ \frac{d \sigma (pN \to cY)}{dx_E}&=& \frac{\pi 
\alpha^2_s (\mu_R)}{m_c^2} \times \\ 
&& \hskip-8em \sum_{ij} \Biggr[ \int^1_0 dx_1 x^{\gamma}_1 
f^1_i(x_1,\mu_F^2)\ \Biggr] \Biggr[\int^1_{\epsilon} dx_2 f^2_j(x_2,\mu_F^2) 
\zeta_{ij}\! \left(\frac{\epsilon}{x_2},\mu_R, \mu_F\right)\Biggr] ,  
\nonumber
\end{eqnarray} 
where the functions $\zeta_{ij}$ are defined as 
\begin{equation} 
\zeta_{ij}(v,\mu_R,\mu_F) = v \int^{1-v}_0 d \tau_2 ~\tau_2^{\gamma + 1}~ 
\int_0^{1-v-\tau_2} d\tau_x~h_{ij} (\tau_x, \tau_2, 4 v \tau_2, \mu_R, 
\mu_F)~,  \label{eq:24} 
\end{equation} 
and the argument $v$ is $v \equiv \epsilon/ x_2$ (to rewrite the integration in
$\tau_2$ we noticed that $\rho/ 4 \tau_2 = v$). The functions $h_{ij}$ are
given by $h_{ij} (\tau_x, \tau_2, \rho, \mu_R, \mu_F) = h^{(0)}_{ij}
(\tau_2,\rho) \delta (\tau_x) + O (\alpha^2_s)$. We will take only gluons as
partons from now on, thus $f^1_i(x,\mu_F^2)=f^2_j(x,\mu_F^2)=g(x,\mu_F^2)$.
 
The function $\zeta_{gg}$, using $h_{gg}$ at the Born level, is shown in
Fig.~6a for $\gamma = 1.7$ and 2 (corresponding to the spectral indices $
\gamma + 1$ of the primary flux above and below the knee). In the same figure
we see that the maximum of $\zeta_{gg} (v)$ is at $v \simeq 0.1$, namely $x_2
\simeq 10~\epsilon$. However, given that $g (x_2,\mu_F^2)$ is a sharply
increasing function with decreasing $x_2$ (i.e. for increasing $v$ at fixed
$E_c$), the maximum of the product $g (x_2,\mu_F^2) \zeta_{gg} (v)$ is always
to the right of the maximum of $\zeta_{gg} (v)$, at $v > 0.1$.  Therefore, the
integral in $x_2$ in Eq.~(28) is dominated by the values of $
x_2$ of order $\epsilon$, namely 
\begin{equation} 
x_2 \simeq \epsilon \simeq \frac{{\rm GeV}}{2E_c}~.  \label{eq:25} 
\end{equation} 
 
Returning to Eq.~(28), the integral in $x_1$ shows that large values of $x_1$
will be dominant since $x_1^{\gamma} g (x_1) \to x_1^{\gamma - \lambda - 1}$
for small $x$, where the exponent is positive, since $\gamma = 1.7$ or 2, while
$0 \lesssim \lambda \lesssim 0.5$ (thus $\gamma - \lambda - 1> 0$). To see more
precisely what range of $ x_1$ dominates the integral, it is necessary to prove
two statements. The first is that $\tau_2 \equiv x_E/x_1 < 1$, due to
kinematical constrains, therefore $x_1 > x_E$. The second is that the
characteristic value of $x_E $ is $0.1$, namely that the $c$-quark is mainly
produced with $0.1$ of the
proton energy 
\begin{equation} 
E_c = {\cal O}( 0.1~ E).  \label{eq:26} 
\end{equation} 
 
With respect to the kinematical limit on $\tau_2$, as we already mentioned, $
\tau_2 \equiv x_E/x_1 \leq 1-v $, and we obtained as a kinematical constraint
that $\epsilon \leq v=\epsilon / x_2 \leq 1$ (since $x_2$ goes from $\epsilon$
to 1). Thus, $\tau_2 \leq 1- \epsilon < 1$, since $\epsilon$ is always larger
than zero. Another way of obtaining this bound is the following. Since the
partonic processes involved are $gg \to c {\bar{c}}$ or $gg \to c {\bar{c}} g$,
then $\sqrt{{\hat{s}}} \geq 2 ({\hat{E}}_c)_{{\rm max }}$ and due to $m_c \not=
0$, (${\hat{k}}_{\/\/})_{{\rm max}} < ({\hat{E}} _c)_{{\rm max}}$, therefore
$\tau_2 < 2 ({\hat{E}}_c)_{{\rm max}} / \sqrt{{ \hat{s}}}\leq 1$.
 
That in fact $E_c = {\cal O}( 0.1~ E)$ is clearly demonstrated in Fig.~6b,
which shows the function $x_E^{\gamma} (d \sigma/dx_E)$ normalized by the total
c-production cross section. Thus we have proven that the dominant range of
$x_1$ in Eq.~(28) is $x_1 \gtrsim {\cal O}(0.1 E)$ and also, combining
together Eq. (\ref{eq:25}) and Eq. (\ref{eq:26}), our statement in Eq.
(\ref{eq:12}) about $x_2$.
 
Even if we have not yet included gluon shadowing in our calculations, we want
to point out that this effect might only be important for the target gluon
(given that $x_2$ is very small) but it is not important for the gluons in the
projectile (given that $x_1 \gtrsim 0.1$). This means that the uncertainties on
the composition of cosmic rays will not affect the results through shadowing
effects.
 
As a summary of our arguments we can say that, due to the incoming flux being
rapidly falling with increasing energy of the primary, only the charm quarks
produced with a large fraction of the incoming energy, $E_c \simeq 0.1 E$,
count in the charm flux at production, and those highly energetic $c $ quarks
come from projectile partons carrying a large fraction of the incoming momentum
$x_1 \gtrsim x_E \simeq 0.1$. On the other hand, because typical partonic
center of mass energies $\sqrt{\hat s}$ are close to the $c \bar c$ threshold,
$2m_c \simeq 2$ GeV (since the cross section decreases steeply with increasing
$\sqrt{\hat s}$), while the total center of mass energy squared is $s = 2 m_p
E$ (with $m_p$ the proton mass, $m_p \simeq 1$ GeV), the product $x_1 x_2
\equiv \hat s/s = 4 m_c^2/(2 m_p E) \simeq$ GeV/$E $. This shows that $x_2
\simeq $ (GeV/$E x_1) \simeq$ GeV/$0.1 E $.
 
We now derive the dependence on $\lambda$ of the muon
and neutrino fluxes for a simple power law primary flux.

We can explain first the dependence on $\lambda$ of the spectral index of
$d \phi_c/dX$ at large energies, and then, using this result, the dependence on
$\lambda$ of the spectral indices of atmospheric muons and neutrinos. To start
with, we notice that the integral in Eq.~(28) depends on the charm energy $E_c$
only through the presence of the parameter $\epsilon$ in the integration on
$x_2$. To approximately perform this integration at large energies, let us
replace $g (x_2) \simeq x_2^{-\lambda - 1}$ in Eq.~(28) and take $\zeta (
\epsilon / x_2) \simeq \zeta_{{\rm max}}$ (namely develop $\zeta$ in powers of
$v= \epsilon/x_2$ and keep only the constant
term) then 
\begin{equation} 
\int^1_{\epsilon} dx_2~ g (x_2) ~\zeta \Biggr(\frac{\epsilon} {x_2} \Biggr) 
\simeq \zeta_{{\rm max}} \int^1_{\epsilon} dx_2~ x_2^{-\lambda - 1}~. 
\label{eq:27} 
\end{equation} 
Since $\epsilon \ll 1$, this integral is well approximated by $\zeta_{{\rm
    max}} \epsilon^{-\lambda}/ \lambda$, for all $\lambda \not= 0$. Better
approximations to the function $\zeta$ give similar results. For example,
approximating the function $\zeta$ by two power laws, one above and another
below the maximum, which is at about $x_2 =5 \epsilon$ ($\zeta = \zeta_{{\rm
    max}}(x_2/5\epsilon)^{2.1}$ for $x_2$ between $\epsilon$ and $5 \epsilon$
and $\zeta = \zeta_{{\rm max}}(5\epsilon/x_2)^{0.4}$ for $x_2$ between $5
\epsilon$ and $1$), the integral in Eq.(\ref{eq:27}) becomes $\zeta_{{\rm max
    }} (5\epsilon)^{-\lambda}/(0.9 +1.7 \lambda - {\lambda}^2)$. Thus the
essential dependence of $\epsilon^{-\lambda}$ is maintained. Recalling that $
\epsilon = m^2_c/(2~m_p ~E_c)$, Eq. (\ref{eq:16}) is proportional to $
E^{\lambda}_c$, and the same is true for Eq. (\ref{eq:27}), therefore
\begin{equation} 
\frac{d \phi_c}{dX}(E_c, X) \sim E_c^{-\gamma - 1 + \lambda} \ \ . 
\label{eq:28} 
\end{equation} 
 
The charm production function $d\phi_c (E_c, X)/dX$, calculated numerically, is
shown in Fig.~7a for a typical $X= 57.12~{\rm g/cm^2}$ ($h$ = 20 km).  We are
using here the PDF MRS R1 with the three related values of $\lambda=0$,
$\lambda$(R1), 0.5. We clearly see here that the slope at $E_c \gtrsim 10^5
{\rm GeV}$ depends on the extrapolation of the gluon PDF at $x <10^{-5}$. This
is one order of magnitude lower in energy than in Fig.~1 for the total cross
section. This reflects the fact mentioned above that the characteristic charm
energy is $E_c = {\cal O}(0.1E)$.  Fig.~7b shows that, as predicted
analytically, the slopes (the negative of the spectral index in our notation)
of the charm fluxes at production depend almost linearly on $ \lambda$. In fact,
in Fig.~7b, we can see that the logarithmic slopes of the $\lambda = 0$ and
$\lambda = 0.5$ fluxes differ precisely by 0.5, above 5 10$^6~{\rm GeV}$
(namely, above the knee) to about $10^9~{\rm GeV}$ (the maximum energy at which
our fluxes are reliable, given that we take $10^{11}$ GeV as the maximum
incoming proton energy $E$). In fact, the slope of the $\lambda = 0$ flux in
that interval is about -3.1 to -3.2, while that of the $\lambda = 0.5 $ flux is
about -2.6 to -2.7. Above the knee, the primary spectrum goes as $ E^{\delta}$
with $\delta \simeq (-\gamma-1-0.1)= - 3.1$, where we have also included the
$0.1$ contribution coming from the $E$-dependence of $\Lambda_N$ (see footnote
in previous discussion), thus the charm spectrum, (in the energy range $10^7
~{\rm GeV} \lesssim E_c \lesssim 10^9~{\rm GeV}$) goes approximately as $
E_c^{\delta +\lambda}$ as expected from Eq. (\ref{eq:28}).
 
Using the definition of the leptonic fluxes in terms of the charm spectrum at
production $d\phi_c/dX$, we can now find the dependence of the spectral index
of muon and neutrino fluxes with $\lambda$. For example, the differential flux
$\phi_{\mu}$ of muons with energy $E_\mu$ ($\mu$ stands here for $\mu^+$ or
$\mu^-$) is
\begin{eqnarray}
\phi_\mu(E_\mu) = 2 \int_{X_0}^{\infty}dX
 \int_{E_\mu}^{\infty} dE_c
   {d \phi_c(E_c, X) \over d X}
\left [ {dN_\mu (c \to \mu; E_c, E_\mu, X) \over d E_{\mu} } \right]
\label{eq:29}
\end{eqnarray}
($\phi_{\mu}$ has, thus, units of $[1/$ cm$^2$ s sr GeV]). Here the factor of 2
accounts for the muons produced by ${\bar c}$ and the last square bracket is
the number of muons of energy $E_\mu$ produced at sea level by the cascades,
each cascade initiated by a $c$ quark of energy $E_c$ at a depth $X$.
 
Our results above indicate that we can write the atmospheric charm spectrum at
production as (see Eq.(\ref{eq:28})) ${d \phi_c}(E_c, X)/ dX \simeq F(X)
E_c^{-\gamma - 1 + \lambda}$ with $F(X)$ a function independent of energy.
Replacing this form for ${d \phi_c}(E_c, X)/ dX $ in Eq. (\ref{eq:29}) and
multiplying and dividing by $E_{\mu}^{-\gamma - 1 + \lambda}$ we can write $
\phi_{\mu}$ as
\begin{eqnarray}
\phi_\mu(E_\mu) =2E_{\mu}^{-\gamma-1+\lambda}\int_{X_0}^{\infty}dXF(X)\int_{E_\mu}^{\infty}dE_c{\left ({E_c \over E_{\mu}}\right )}^{-\gamma - 1 + \lambda}
\left [ {dN_\mu (c \to \mu; E_c, E_\mu, X) \over d E_{\mu} } \right]
\label{eq:30}.
\end{eqnarray}

We can argue that in so far as the values of the parent charm quark energy $
E_c$ and the daughter lepton energy $E_{\mu}$ are not very different, the
dependence of the integral on $\lambda$ (and on $\gamma$) should be mild. In
this case, from Eq. (\ref{eq:30}), we find that the spectral index of the muon
(and similarly of the neutrino) flux contains $\lambda$ as a term, i.e.
\begin{equation} 
\phi_{\mu} ( E_{\mu}) \simeq f(E_\mu,\gamma, \lambda) {E_\mu}^{-\gamma - 1 + 
\lambda} \equiv E_{\mu}^{-b_{\mu}(E_\mu, \gamma, \lambda) + \lambda}~, 
\label{eq:31} 
\end{equation} 
where the dependence of the functions $f(E_\mu, \gamma, \lambda)$ and $
b_{\mu}(E_\mu,\gamma,\lambda)$ on $\lambda$ and $\gamma$ should be mild.
This justifies the results shown in Figs.~4 and~5, presented in Sect.~\ref{sect:fluxes},
showing all the spectral indices obtained using all our PDF's.
 
Finally we examine the deviations from linearity of the relation between the
spectral index $\alpha_{\ell}$ and the gluon PDF slope $\lambda$. In Fig.~8a we show directly the relation between $\lambda$ and $\alpha_{\ell}$, using the values coming from our simulation for the MRST case already presented in Fig.~4, but now plotting them for fixed energy $E_{\mu}$. We show two examples, for $E_{\mu} = 1~PeV,~10~PeV$, where our points indicate a good agreement with the linear relation between $\alpha_{\ell}$ and $\lambda$ of Eq.~(\ref{eq:11}). 

The mild dependence on $\lambda $ of the functions $b_{\ell}(\lambda) = \alpha_{\ell}+\lambda$ can be seen in
Fig.~8b, where we show the percentage difference $ [b_{\ell}(\lambda)-b_{\ell}(0)]/b_{\ell}(0) $ for
the different values of $\lambda =0$--0.5 with the MRST PDF.  It is evident
that, in the range where our theoretical arguments are applicable (for $E_{\mu}
\gtrsim 10^{6}~{\rm GeV}$) the $b_{\ell}(\lambda)$ functions differ only by $2-3\%$
for different $\lambda $ values, namely they are almost independent of $\lambda
$, given one particular PDF.  This analysis confirms the validity of
Eq.~(\ref{eq:11}), which leads to the possibility of obtaining information on $
\lambda $ at small parton fractions $x$ not reachable in experiments, through
the measurement of the fluxes. We will study this possibility in more detail in
a future paper \cite{GGV3}.
 
\section{Conclusions} 
 
\label{sect:conclusions} 
 
The actual next-to-leading order perturbative QCD calculations of charm
production cross sections, together with a full simulation of the atmospheric
cascades, were used to obtain the vertical prompt fluxes of neutrinos and
muons.
 
We have analyzed the dependence of the atmospheric fluxes on the extrapolation
of the gluon PDF at very low $x$, which is related to the value of the
parameter $\lambda $. This was done using four different sets of PDF's: MRS R1,
MRS R2, CTEQ 4M and MRST, with variable $\lambda $ in the range 0--0.5.
 
The charm production cross sections and the final lepton fluxes depend
critically on $\lambda$ for leptonic energies $E_l \gtrsim 10^5~{\rm GeV}$,
which correspond to $x \lesssim 10^{-5}~{\rm GeV}$. We found that the fluxes
vary up to almost two orders of magnitude at the highest energy considered,
$10^9~{\rm GeV}$, for the different $\lambda$'s in the allowed interval; on the
contrary, for fixed $\lambda$, the results don't depend much on the choice of
the PDF.
 
For the lowest values of $\lambda$ ($\lambda \simeq 0-0.1$) our fluxes are very
close to those of TIG \cite{TIG}, confirming that the very low flux prediction
is mostly due to a low value of $\lambda$ ($\lambda_{\rm TIG} \simeq 0.08$).
For higher values of $\lambda$ ($\lambda \simeq 0.2-0.5$) our results are in
the bulk of previous predictions and, in particular, for $ \lambda \simeq 0.3$
they are very close to a recent semi-analytical calculation \cite{PRS} done with
a similar value of $\lambda$.
 
We have also considered the dependence of the spectral index of the final
fluxes on the parameters of the model. From both, computer simulations and
analytical considerations, we find that the spectral index $\alpha_{\ell}$ of
atmospheric leptonic fluxes depends linearly on $\lambda$ as in Eq.~(\ref
{eq:11}).
 
This suggests the possibility of obtaining bounds on $\lambda$ in ``neutrino
telescopes" for small values of $x$ not reachable in colliders, if the spectral
index of leptonic atmospheric fluxes could be determined by these telescopes.
We will investigate this possibility in detail in the future \cite{GGV3}.
 

\acknowledgments
The authors would like to thank the Aspen Center For Physics, where
this work was initiated, for hospitality, and M. Mangano and P. Nason for the
MNR program and helpful discussions. This research was supported in part by the
US Department of Energy under grant DE-FG03-91ER40662 Task C.

\newpage 
 
\section*{FIGURE CAPTIONS} 
 
\begin{description} 
\item[Fig.~1] Total cross section for charm production
  $\sigma_{c\overline{c}}$, up to NLO, for our different PDF's and $\lambda$
  values, compared to that used by TIG \cite{TIG}. Top panel: MRS
  R1-R2 and CTEQ 4M; bottom panel: MRST (cross sections increase with
  $\lambda$).
 
\item[Fig.~2] Prompt muons: $E^{3}$-weighted vertical fluxes at NLO,
  compared to the TIG \cite{TIG} conventional and prompt fluxes (dotted lines).
  We show results using the four PDF's MRS R1, MRS R2, CTEQ 4M and MRST.
 
\item[Fig.~3] Prompt muon-neutrinos: $E^{3}$-weighted vertical
  fluxes at NLO, compared to the TIG \cite{TIG} conventional and prompt fluxes
  (dotted lines).  We show results using the four PDF's MRS R1, MRS R2, CTEQ 4M
  and MRST. 

\item[Fig.~4] Prompt muons: spectral index of the NLO vertical fluxes for the
  four PDF's MRS R1, MRS R2, CTEQ 4M and MRST.
 
\item[Fig.~5] Dependence of prompt fluxes and their spectral index on the PDF
  at fixed $\lambda$: left side $\lambda=0$, right side $\lambda=0.5$.
 
\item[Fig.~6] (a) The function $\zeta_{gg}(v)$ at the Born level for $\gamma =     0,~1.7$ (below the knee) and $\gamma = 2$ (above the knee). (b) Flux-weighted charm production spectra $x_E^\gamma
  {\frac{1}{\sigma}} {\frac{d\sigma}{dx_E}}$ at several beam energies (using
  MRS R1, $\lambda$(R1)).
 
\item[Fig.~7] (a) NLO charm production function $E_{c}^3 d\phi_c(E_c,X)/dX$ (PDF MRS R1); (b) its
  spectral index $-\partial\ln [\partial \phi_c(E_c,X)/\partial X]/\partial\ln E_c$.
  These results are for a height $h = 20$ km, corresponding to a vertical depth
  $X = 57.12~{\rm g/cm^2}$ (similar results are obtained for other heights).
   
\item[Fig.~8] (a) Relation between the slope $\lambda$ of the gluon PDF and the
  muon spectral index $\alpha_\mu$ at fixed muon energy. (b) Non-linearities in
  this relation. Here $b_{\ell}(\lambda) = \alpha_{\ell}(\lambda)+\lambda$ and we use the
  MRST PDF.

\end{description} 
 
\newpage
\begin{figure}[t]
\epsfig{file=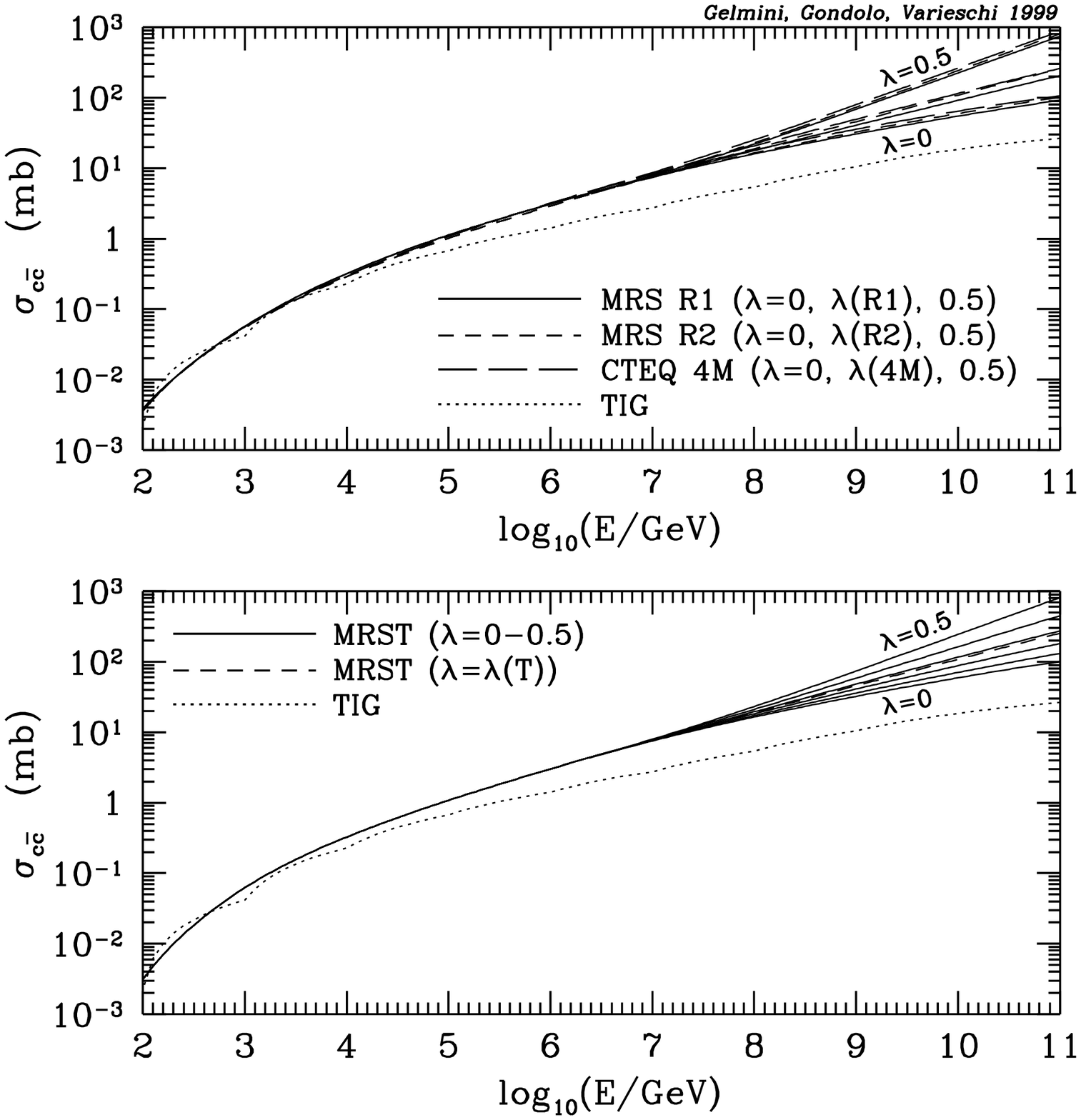,width=\textwidth}
\caption{~}
\end{figure}

\newpage
\begin{figure}[t]
\epsfig{file=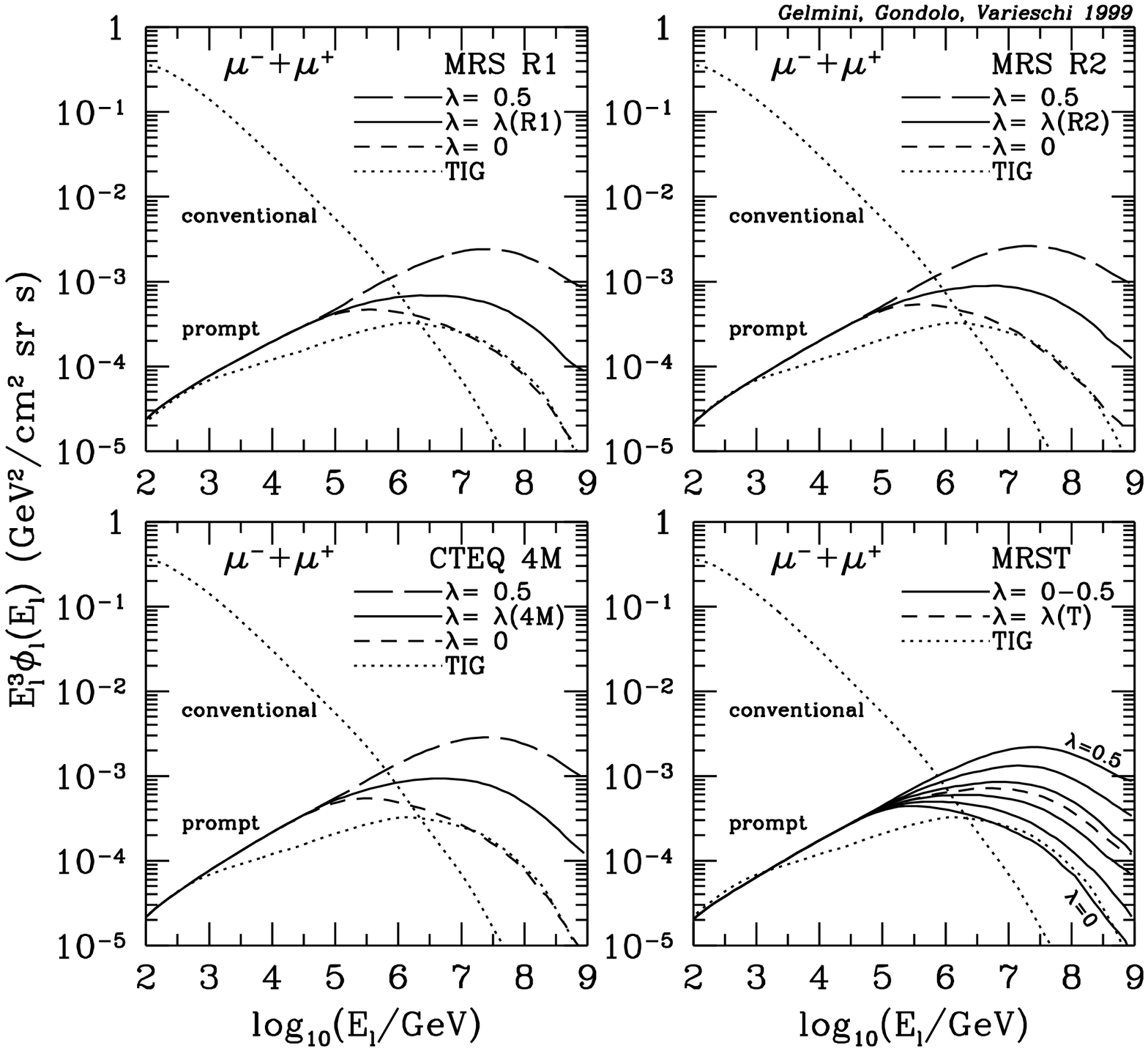,width=\textwidth}
\caption{~}
\end{figure}

\newpage
\begin{figure}[t]
\epsfig{file=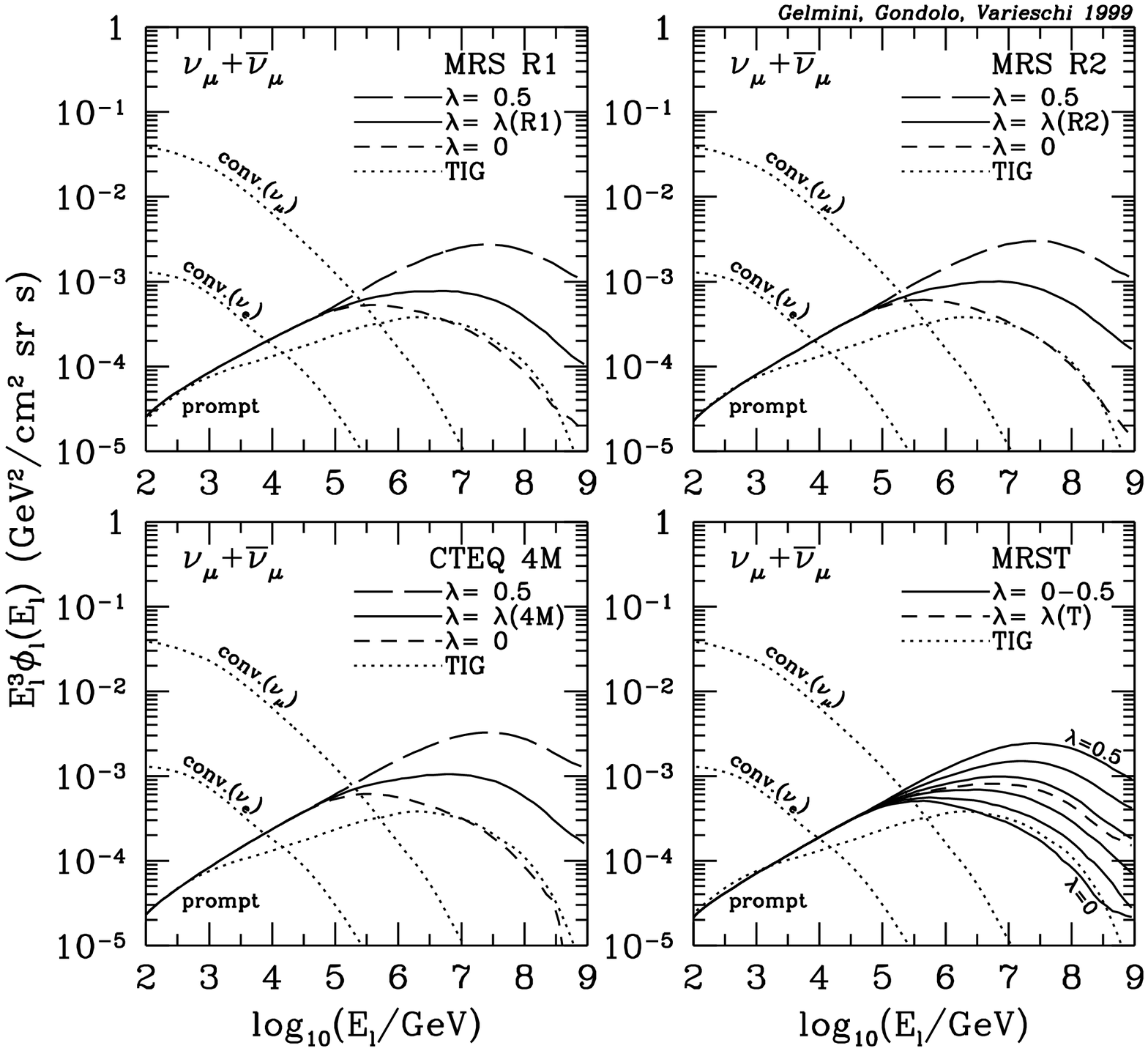,width=\textwidth}
\caption{~}
\end{figure}

\newpage
\begin{figure}[t]
\epsfig{file=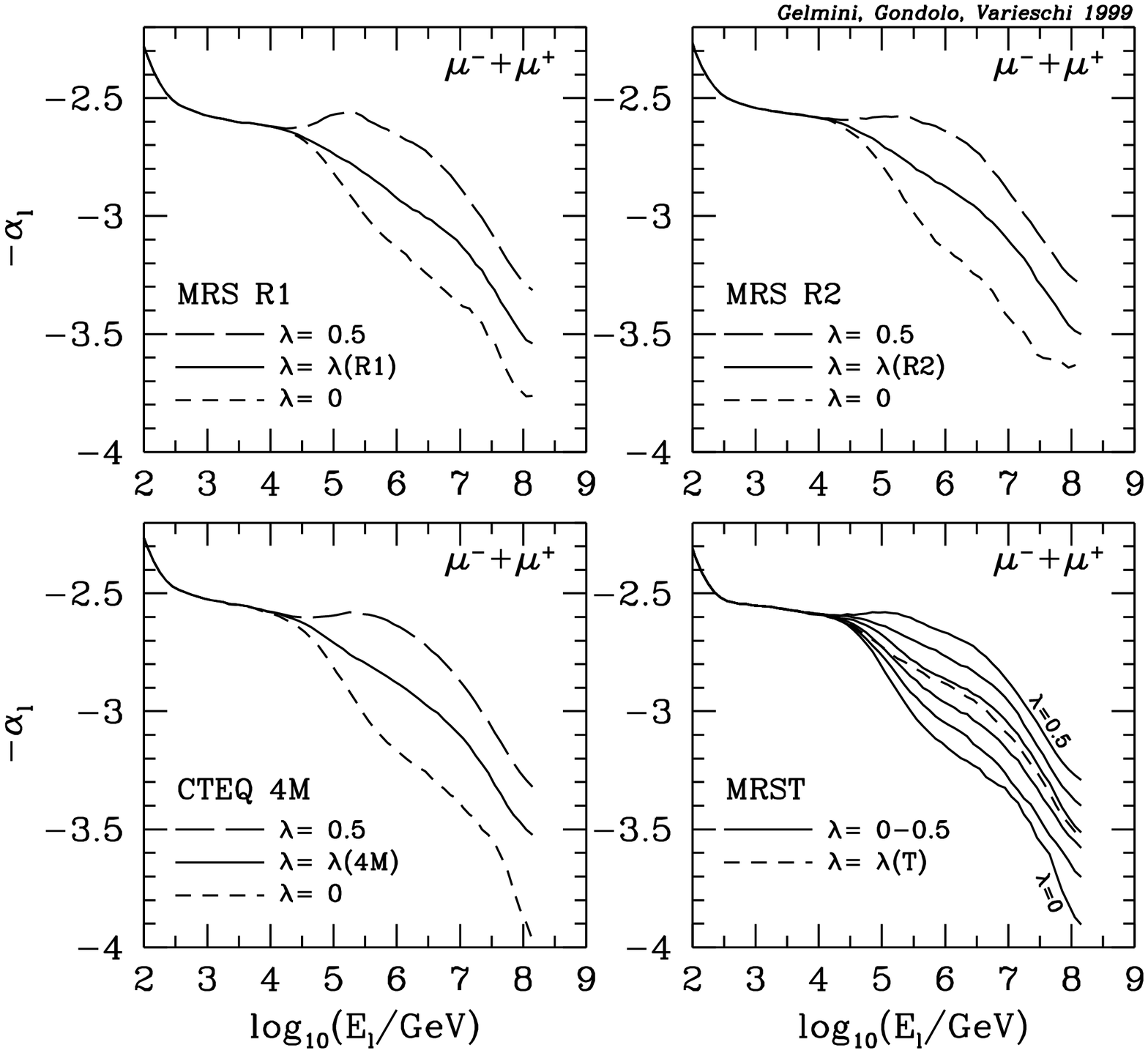,width=\textwidth}
\caption{~}
\end{figure}

\newpage
\begin{figure}[t]
\epsfig{file=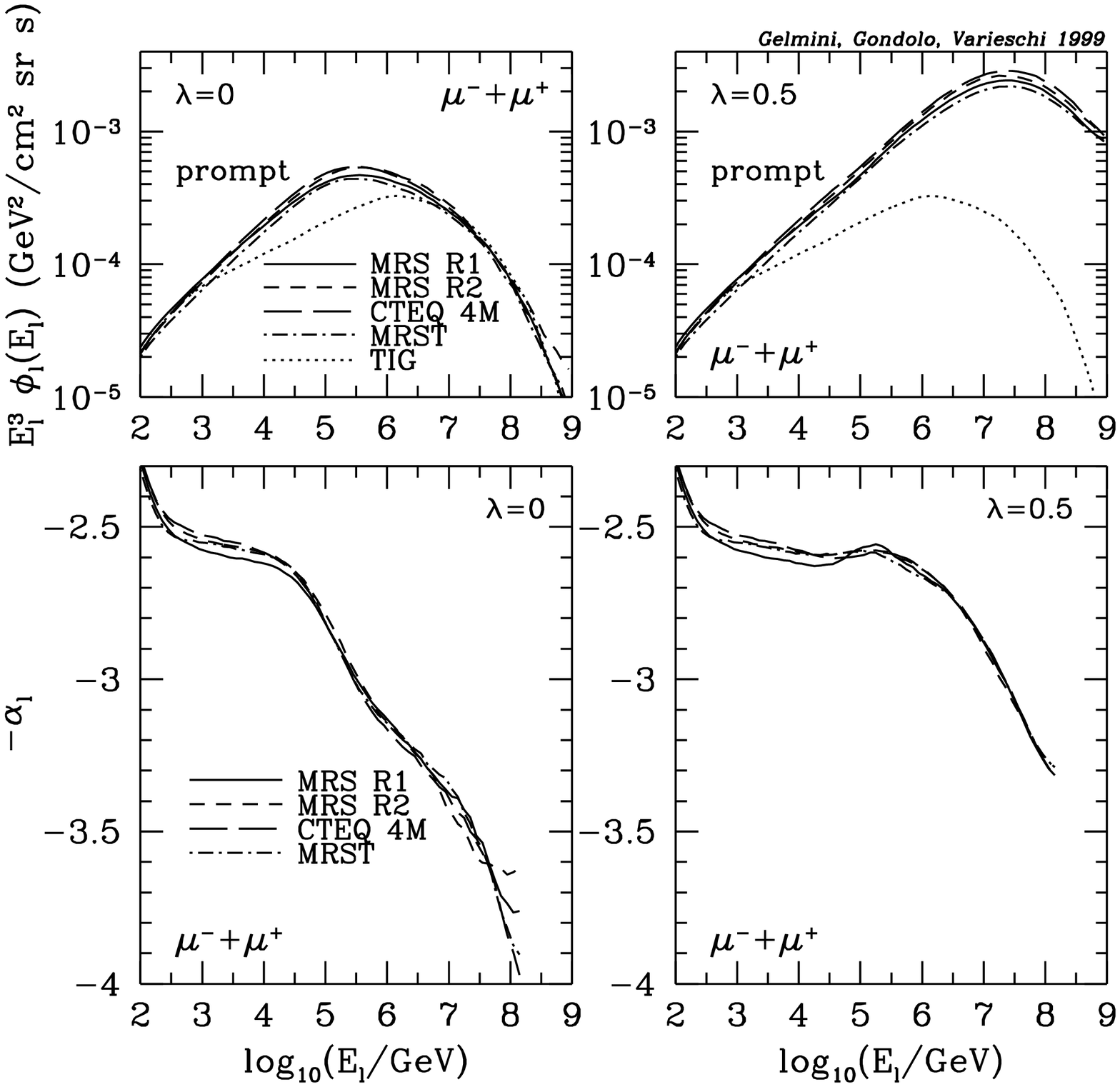,width=\textwidth}
\caption{~}
\end{figure}

\newpage
\begin{figure}[t]
\epsfig{file=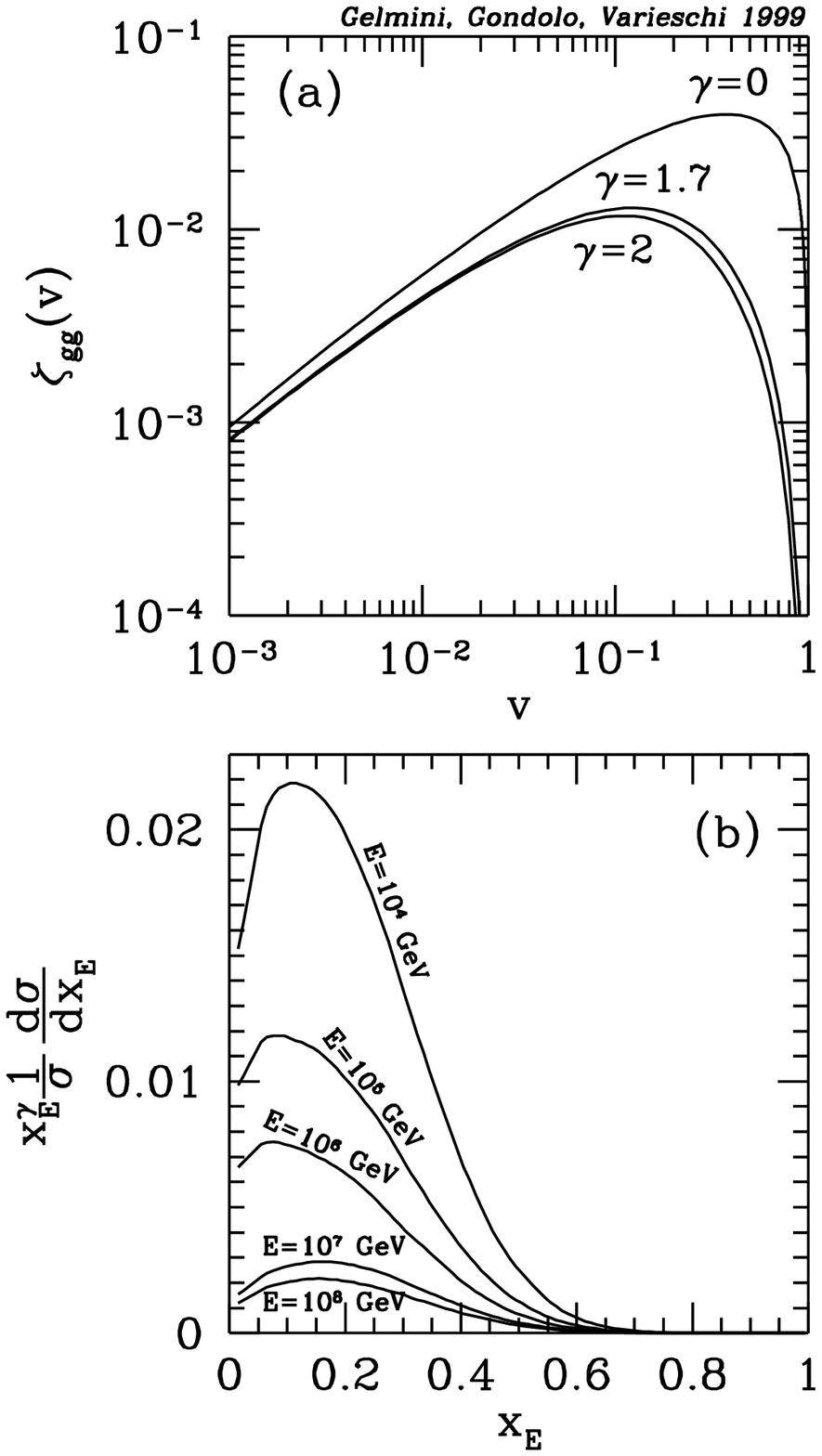,width=\textwidth}
\caption{~}
\end{figure}

\newpage
\begin{figure}[t]
\begin{center}
\epsfig{file=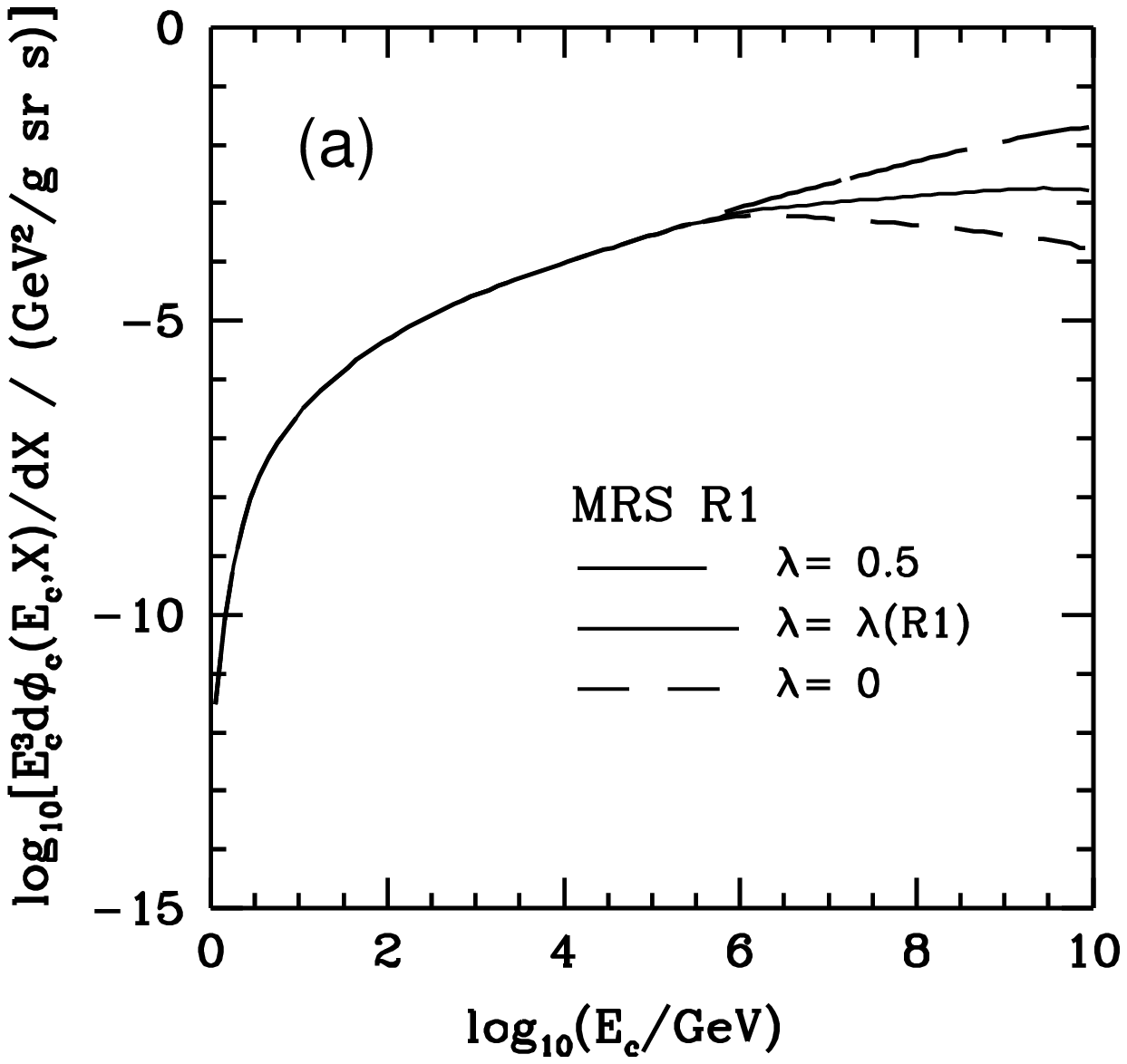,width=0.6\textwidth}
\epsfig{file=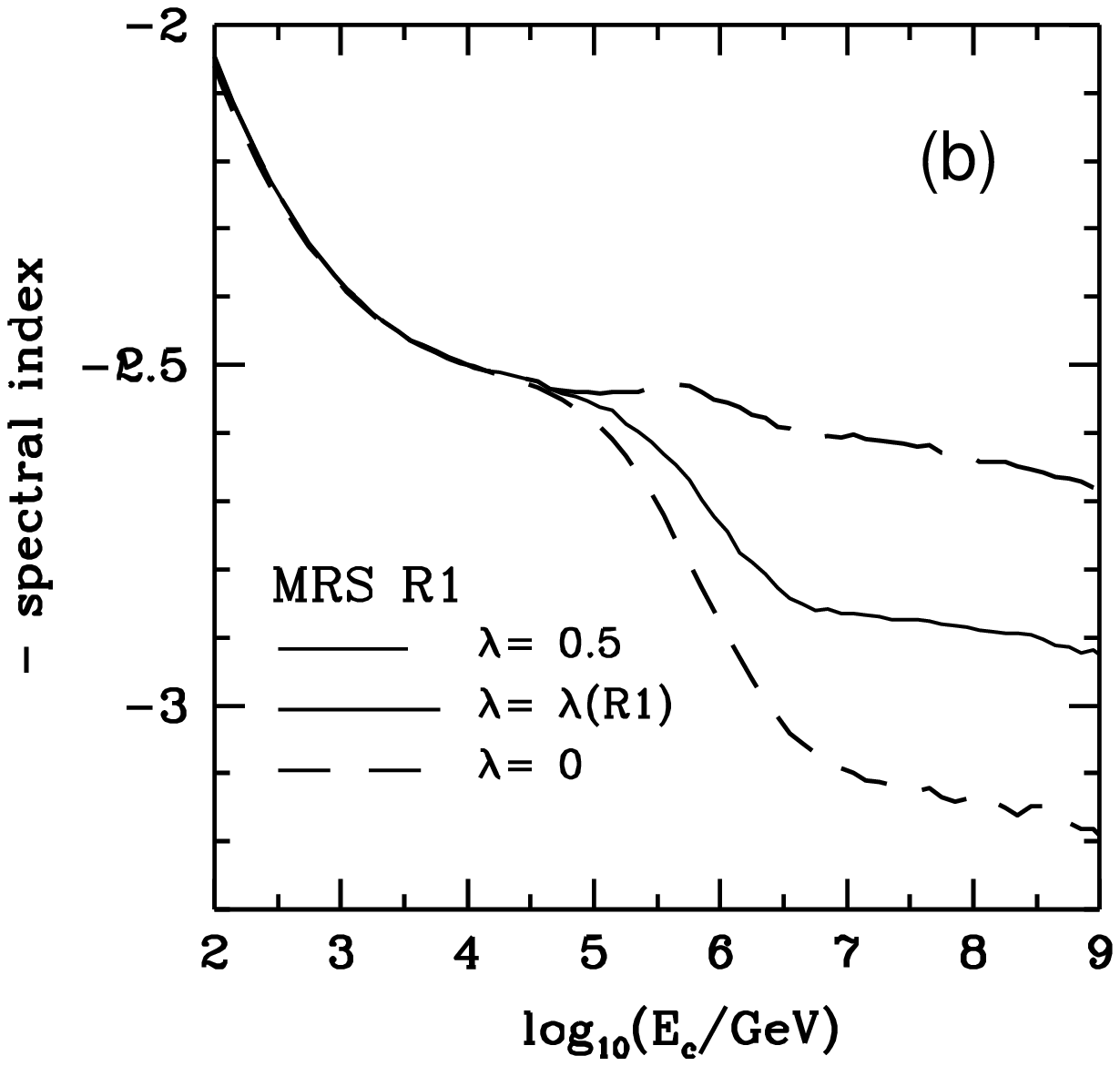,width=0.6\textwidth}
\end{center}
\caption{~}
\end{figure}

\newpage
\begin{figure}[t]
\epsfig{file=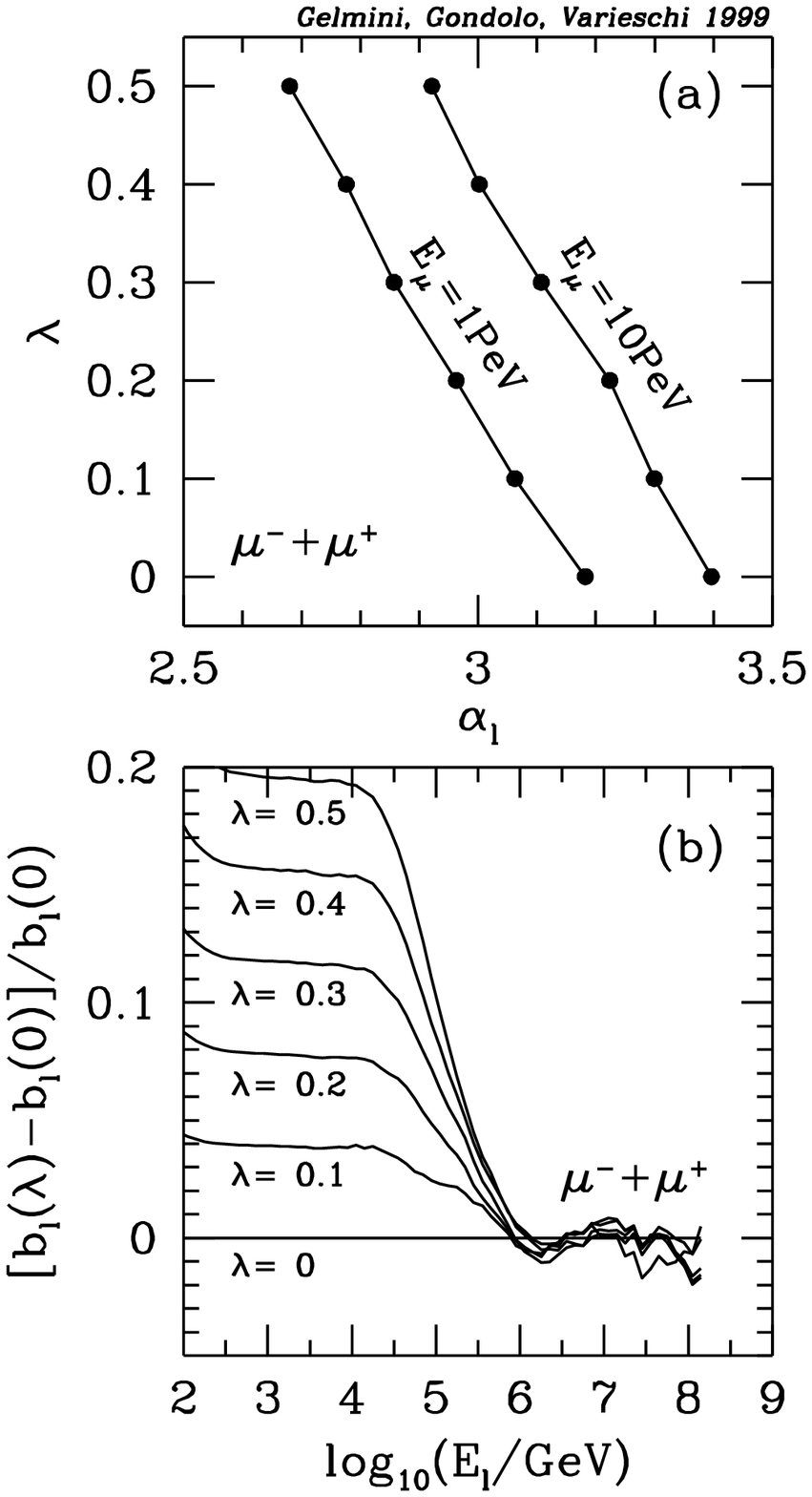,width=\textwidth}
\caption{~}
\end{figure}

\end{document}